\documentclass[preprint,aps,showpacs]{revtex4}

\begin{document}

\title{Nonlinear wave interactions in quantum magnetoplasmas }

\author{P. K. Shukla\footnote{ Also at: Centre for Nonlinear Physics, Department of Physics, 
Ume\aa ~University, SE-90187 Ume\aa,~ Sweden; Max-Planck Institut f\"{u}r 
extraterrestrische Physik, D-85741 Garching, Germany; GoLP/Instituto Superior T\'{e}cnico,
1049-001 Lisbon, Portugal; CCLRC Centre for Fundamental Physics, Rutherford
Appleton Laboratory, Chilton, Didcot, Oxon 0X11 0QX, UK; SUPA Department of
Physics, University of Strathclyde, Glasgow G 40NG, UK}\email{ps@tp4.rub.de}
and S. Ali\footnote{Also at: Department of Physics,
Government College University, Lahore 54000, Pakistan}\email{shahid\_gc@yahoo.com}}

\affiliation{Institut f\"{u}r Theoretische Physik IV and Centre for Plasma Science and\\
Astrophysics, Fakult\"{a}t f\"{u}r Physik und Astronomie,\\
Ruhr--Universit\"{a}t Bochum, D--44780 Bochum, Germany}

\author{L. Stenflo\email{lennart.stenflo@physics.umu.se} and M. Marklund\footnote{Also at: CCLRC Centre for Fundamental Physics, Rutherford
Appleton Laboratory, Chilton, Didcot, Oxon 0X11 0QX, UK}\email{mattias.marklund@physics.umu.se}}

\affiliation{Centre for Nonlinear Physics, Department of Physics, Ume\aa\ University, 
SE-90187 Ume\aa , Sweden}





\begin{abstract}
Nonlinear interactions involving electrostatic upper-hybrid (UH),
ion-cyclotron (IC), lower-hybrid (LH), and Alfv\'{e}n waves 
in quantum magnetoplasmas are considered. For this purpose, the quantum hydrodynamical
equations are used to derive the governing equations for nonlinearly coupled
UH, IC, LH, and Alfv\'{e}n waves. The equations are then Fourier analyzed to
obtain nonlinear dispersion relations, which admit both decay and
modulational instabilities of the UH waves at quantum scales. 
The growth rates of the instabilities are presented. 
They can be useful in applications
of our work to diagnostics in laboratory and astrophysical settings. 
\end{abstract}


\maketitle

\section{Introduction}

Quantum plasma physics is a new and rapidly emerging subfield of plasma physics.
It has received a great deal of attention due to its wide range of applications \cite
{1,2,3,4,5}. Quantum plasmas can be composed
of the electrons, positrons, holes, and ions. They are characterized by
low temperatures and high particle number densities. Quantum plasmas and collective effects 
play an important role in 
microelectronic components \cite{1}, dense astrophysical systems
(in particular white dwarf and neutron star environments) \cite{2}, intense laser-matter
experiments \cite{3}, and nonlinear quantum optics \cite{4,5}. It is well
known that when the thermal de Broglie wavelength of the charged particles
is equal to or larger than the average inter-particle distance $d=n^{-1/3}$, where
$n$ is a typical plasma density,   
the quantum mechanical
effects play a significant role in the behaviour of the charged particles. 
There are two well-known mathematical formulations, 
the Wigner-Poisson and the Schr\"{o}dinger-Poisson approaches, that have 
been widely used to describe the statistical and hydrodynamic behavior of the 
plasma particles at quantum scales in quantum plasmas. These
formulations are the quantum analogues of the kinetic and the fluid models in
classical plasma physics. Manfredi \cite{6} has 
studied these approaches, taking into account the quantum effects in a
collisionless quantum plasma. In particular, the quantum
hydrodynamic model (QHD) has attracted much interest in studies of the
negative differential resistance \cite{7} in the tunnelling diode. Several
collective processes\cite{8,9,10,11,12,13,14} have been analyzed both
analytically and numerically in plasmas with quantum corrections.

Haas {\it et al.} \cite{15} studied a quantum multi-stream model for
one- and two-stream plasma instabilities, presented a new purely quantum
branch, and investigated the stationary states of the nonlinear
Schr\"{o}dinger-Poisson system. Anderson {\it et al.}\cite{16} 
used a Wigner-Poisson formulation showing that Landau-like damping 
due to phase noise can suppress the instabilities.
Furthermore, a detailed study of the linear and nonlinear properties of ion
acoustic waves (IAW) in an unmagnetized quantum plasma has been presented 
by Haas {\it et al.} \cite{17}. For this purpose, they employed the QHD equations
containing a non-dimensional quantum parameter $H.$ The latter is
the ratio between the plasmon and thermal energies. For a weakly nonlinear
quantum IAW, a modified Korteweg-de Vries (KdV) equation was analyzed
for $H\rightarrow 2$, $H$ $<2$ and $H>2$, connected with a shock wave, as well as
bright and dark solitons, respectively. Finally, they also observed a
coherent, periodic pattern for a fully nonlinear IAW in a quantum
plasma. Such a pattern cannot exist in classical plasmas. The formation and 
dynamics of dark solitons and vortices in quantum electron plasmas has also been
reported by Shukla and Eliasson \cite{r17a}.

Recently, Haas \cite{18} extended the QHD equations for quantum
magnetoplasmas and presented a magnetohydrodynamic model by using the
Wigner-Poisson system. He pointed out the importance of the external magnetic field, 
by establishing the conditions for equilibrium in ideal quantum
magnetohydrodynamics. Garcia {\it et al. }\cite{22} derived the 
quantum Zakharov equations by considering a one-dimensional quantum system 
composed of electrons and singly charged ions. They also investigated 
the decay and four-wave instabilities for the nonlinear coupling between 
high-frequency Langmuir waves and low-frequency IAWs. Marklund 
\cite{23} considered the statistical aspect and solved the Zakharov system
at quantum scales, and analyzed the modulational instability both analytically
and numerically. Recently, Shukla and Stenflo \cite{24} investigated 
parametric and modulational instabilities due to the interaction of 
large amplitude electromagnetic waves and low-frequency electron and ion
plasma waves in quantum plasmas. Drift modes in quantum plasmas \cite{shukla-stenflo1}, as well as 
new modes in quantum dusty plasmas \cite{stenflo-etal,shukla-stenflo2}, have also
been considered. 

In the past, Yu and Shukla \cite{19} studied the nonlinear coupling of
UH waves with low-frequency IC waves and obtained near-sonic UH cusped
envelope solitons in a classical magnetoplasma. The nonlinear dispersion 
relations \cite{20} were also derived for three wave decay interactions
and modulational instabilities due to nonlinear interactions of 
mode-converted electron Bernstein and low-frequency waves, such as IAWs, 
electron-acoustic waves (EAWs), IC waves, quasimodes, magnetosonic waves, 
and Alfv\'en waves.  Murtaza and Shukla \cite{21} illustrated the nonlinear 
generation of electromagnetic waves by UH waves in a uniform magnetoplasma.
Kaufman and Stenflo \cite{kaufman-stenflo} considered the interaction between 
UH waves and magnetosonic modes, and showed that UH solitons
could be formed. 

In the present paper, we consider the nonlinear interactions between UH waves,
IC waves, LH waves, and Alfv\'{e}n waves in a quantum magnetoplasma, by using 
the one-dimensional QHD equations. Both decay and modulational instabilities will be 
analyzed in quantum settings. The manuscript is organized in the following fashion: 
In Sec. II, we derive the governing equations for nonlinearly coupled UH waves, 
IC waves, LH waves, and Alfv\'{e}n waves in quantum plasmas. The coupled equations 
are then space-time Fourier transformed to obtain the dispersion relations. 
The latter admit  a class of parametric instabilities of the UHs. Details 
of the decay and modulational instabilities in quantum plasmas are presented 
in Sec. III. Section IV summarizes our main results.

\section{Nonlinear Dispersion Relations}

In this section, we derive the governing equations and dispersion relations for nonlinearly 
coupled UH, IC, LH, and Alfv\'{e}n waves in a quantum magnetoplasma by using the  
one-dimensional QHD equations \cite{18}.

\subsection{UH waves}

Let us consider the nonlinear propagation of an electrostatic UH wave in a
cold quantum plasma embedded in an external magnetic field $B_0{\bf \hat{z}}$, 
where $B_0$ is the strength of the magnetic field and ${\bf \hat{z}}$ is
the unit vector along the z-axis in a Cartesian coordinates system. The UH
wave electric field is ${\bf E} \approx \hat {\bf x} E_{x0}\exp(i{\bf k}_0\cdot {\bf r%
}-i\omega _0t)$ $+$ complex conjugate, where $\mathbf{k}_0$ is the wave vector and 
$\omega_0$ is the wave frequency.
We then assume that the parallel electric field is small, i.e.\ $E_z\ll E_x$. In the presence of the
electron density fluctuation $n_{e1}$ ($n_{e1} \ll n_{e0}$, where $n_{e0}$ is the unperturbed electron
number density) of the electrostatic IC and LH waves, as well as of the 
magnetic field fluctuation of the Alfv\'{e}n waves, the UH wave dynamics is
here governed by the continuity equation 

\begin{equation}
\frac{\partial n_{e1}}{\partial t}+n_{e0}\frac \partial {\partial x}(1+N_s)%
\text{ }U_{ex}=0\text{ ,}  \label{1}
\end{equation}
the $x$- and $y$-components of the electron momentum equation

\begin{equation}
\frac{\partial U_{ex}}{\partial t}=-\frac e{m_e}E_x - \omega _{ce}\left( 1+%
\frac{B_1}{B_0}\right) U_{ey}+\frac{\hbar ^2}{4m_e^2n_{e0}}\frac \partial {%
\partial x}\nabla ^2n_{e1}\text{ ,}  \label{2}
\end{equation}
 
\begin{equation}
\frac{\partial U_{ey}}{\partial t} = \omega _{ce}\left( 1+\frac{B_1}{B_0}%
\right) U_{ex}\text{ ,}  \label{3}
\end{equation}
and the Poisson equation 

\begin{equation}
\frac{\partial E_x}{\partial x}=-4\pi en_{e1}\text{ },  \label{4}
\end{equation}
where $\omega _{ce}=eB_0/m_ec$ is the
electron gyro frequency, $e$ is the magnitude of the electron charge, $c$ is the speed of light in vacuum,
 $m_e$ is the electron mass, and  
$\hbar $ is the Planck constant divided by $%
2\pi $. Furthermore,   
$\nabla ^2=\partial _x^2+\partial _z^2,$ $N_s=n_{e1}^s/n_{e0}$ is the
relative electron number density perturbation associated with the 
plasma slow motion, and $B_1(\ll B_0)$ is the compressional magnetic field perturbation 
associated with the Alfv\'{e}n wave. In addition, $U_{ex}$ and $U_{ey}$ are the $x$- and 
$y$-components of the perturbed electron fluid velocity associated with the UH wave, 
respectively. The origin of the last term in the right-hand side of Eq. (2) is the quantum
correlation due to the electron density fluctuations \cite{6} in dense quantum plasmas. 
We have also assumed that the electron pressure term is much smaller than the electron 
quantum diffraction term, i.e.,  $\text{ }V_{Fe}^2n_{e1}\ll (\hbar
^2/4m_e^2)\nabla ^2n_{e1}$, where $V_{Fe}$ is the Fermi speed of the
electrons. 

Combining (1)-(4), we obtain 

\begin{equation}
\left[ \frac{\partial ^2}{\partial t^2}+\omega _H^2+2\omega _{ce}^2\left( 
\frac{B_1}{B_0}\right) +N_s\omega _{pe}^2+(1+N_s)\frac{\hbar ^2}{4m_e^2}%
\frac{\partial ^2}{\partial x^2}\nabla ^2\right] \text{ }E_x=0\text{ ,}
\label{5}
\end{equation}
where $\omega _H=\sqrt{\omega _{pe}^2+\omega _{ce}^2}$ is the UH resonance
frequency, and $\omega _{pe}=\sqrt{4\pi n_{e0}e^2/m_e}$ is the electron plasma frequency. 
In the absense of electron density and magnetic field fluctuations, Eq. (5) reduces
to $[\partial_t^2 + \omega_H^2  + (\hbar^2/4m_e^2)\partial_x^2\nabla^2]E_{x0} = 0$,
i.e.\ the pump wave frequency is $\omega _0=\sqrt{\omega _{pe}^2+\omega _{ce}^2+(\hbar
^2/4m_e^2)k_{x0}^2k_0^2\text{ }}$, where $k_0=\sqrt{k_{x0}^2+k_{z0}^2}$ is the
magnitude of the wavevector. As $k_{z0}$ here is much smaller than $k_{x0}$, 
we can write the pump wave frequency as $\omega _0=\sqrt{\omega _{pe}^2+\omega _{ce}^2+(\hbar
^2/4m_e^2)k_0^4\text{ }}$.

\subsection{Electrostatic IC waves}

In the quasi-neutral approximation $\left( n_{e1}^s\approx n_{i1}^s\right)$, 
we now derive the expression for the electrostatic potential associated with
the IC waves in the presence of the UH ponderomotive force. We assume that the
electrons are inertialess, and obtain from the parallel component of the
electron momentum equation 

\begin{equation}
0=-\frac{e^2\omega_H^2}{4m_e\omega _{pe}^4}\frac \partial {\partial z}\left\langle \left|
E_x\right| ^2\right\rangle+e\frac{\partial \phi }{\partial
z}+\frac{\hbar ^2}{4m_e}\frac \partial {\partial z}\nabla ^2N_s  \label{6}
\end{equation}
or 

\begin{equation}
\phi =\frac{e\omega_H^2}{4m_e\omega _{pe}^4}\left\langle \left| E_x\right|
^2\right\rangle - \frac{\hbar ^2}{4m_ee}\nabla ^2N_s
\label{7}
\end{equation}
The first term in the right-side of (6) is the parallel (to ${\bf \hat{z}}$)
component of the ponderomotive potential of the UH waves. The ion dynamics
associated with the electrostatic\ IC waves are governed by the equation of
continuity 

\begin{equation}
\frac{\partial N_s}{\partial t}+\frac \partial {\partial x}U_{ix}=0\text{ },
\label{8}
\end{equation}
and the $x$- and $y$-components of the ion-momentum equation 

\begin{equation}
\frac{\partial U_{ix}}{\partial t}=-\frac e{m_i}\frac{\partial \phi }{%
\partial x} + \omega _{ci}U_{iy}+\frac{\hbar ^2}{4m_i^2}\frac \partial {%
\partial x}\nabla ^2N_s\text{ ,}  \label{9}
\end{equation}
and 
\begin{equation}
\frac{\partial U_{iy}}{\partial t} = -\omega _{ci}U_{ix}\text{ .}  \label{10}
\end{equation}
We have here ignored the ponderomotive force acting on the ions, since it is
smaller (in comparison with the electron ponderomotive force) by the electron 
to ion mass ratio. Furthermore, $U_{ix}$ and $ U_{iy} $ are the $x$-and $y$-components of the 
perturbed ion fluid velocity associated with the plasma slow motion, respectively, 
$\omega _{ci}=eB_0/m_ic$ is the ion gyrofrequency, and $m_i$ is the ion mass. 

Solving (8)-(10), we obtain

\begin{equation}
\left( \frac{\partial ^2}{\partial t^2}+\omega _{ci}^2\right) N_s=\frac e{m_i%
}\frac{\partial ^2\phi }{\partial x^2}.  \label{11}
\end{equation}
Eliminating $\phi $ from (7) and (11), and invoking the quasi-neutrality condition, we then have

\begin{equation}
\left( \frac{\partial ^2}{\partial t^2}+\Omega _{IC}^2\right) N_s=\frac{e^2\omega_H^2}{%
4m_em_i\omega _{pe}^4}\frac{\partial ^2}{\partial x^2}\left\langle \left|
E_x\right| ^2\right\rangle \text{ ,}  \label{12}
\end{equation}
where $\Omega _{IC}=\left[ \omega _{ci}^2+\left( \hbar ^2/4m_em_i\right)
\partial ^2/\partial x^2\nabla ^2\right] ^{1/2}$ is the ion-cyclotron wave 
gyrofrequency including quantum diffraction effects. In deriving Eq. (12), 
we have assumed 
\[
\text{ }\frac{\partial ^2}{\partial t^2}N_s\gg \frac{\hbar ^2}{m_i^2}\frac{%
\partial ^2}{\partial x^2}\nabla ^2N_s\text{ .} 
\]
Equation (12) is the driven (by the UH ponderomotive force) IC wave
equation. In the absence of the UH waves and using $N_s=\hat{N}_s$ exp$%
(-i\Omega t+i{\bf k}\cdot {\bf r})$ in Eq. (12), we obtain the frequency 
$\Omega$ of the IC waves in a quantum magnetoplasma

\begin{equation}
\Omega ^2=\omega _{ci}^2+\frac{\hbar ^2}{4m_em_i}k_x^2k^2\equiv \Omega
_{IC}^2\text{ ,}  \label{13}
\end{equation}
which shows the dispersion due to quantum electron density correlations.
Here, $k=\sqrt{k_x^2+k_z^2}$ is the wavenumber of the electrostatic IC waves. 
By neglecting the quantum diffraction effects ($\hbar \rightarrow
0), $ the dispersion relation of the usual IC wave in a cold magnetoplasma is obtained. 
Equation (5) with $B_1=0$ and Eq. (12) are the desired set for the nonlinearly
coupled electrostatic UH and IC waves in a quantum magnetoplasma.

\subsection{Electrostatic LH waves}

For the electrostatic LH waves, we assume $\omega _{ci}\ll \Omega \ll \omega _{ce}$, 
so that the ions (electrons) are unmagnetized (magnetized). 
The electron dynamics is then governed by the continuity equation,
the momentum equation including the UH ponderomotive
potential and the electron quantum diffraction effects under the
approximation $\Omega \ll \omega _{ce}$. We have, respectively, 

\begin{equation}
\frac{\partial N_s}{\partial t}+\frac \partial {\partial x}U_{ex}=0\text{ },
\label{14}
\end{equation}
and \cite{shukla-stenflo3}
\begin{equation}
{\bf U}_{e\perp }=\frac c{\omega _{ce}B_0}\frac \partial {\partial t}{\bf %
\nabla }_{\perp }\varphi _e+\frac c{B_0}\left( {\bf \hat{z}}\times {\bf %
\nabla }_{\perp }\right) \varphi _e\text{ .}  \label{15}
\end{equation}
Since the second term in the right-hand side of Eq. (15) does not contribute 
to the x-component of the perturbed electron fluid velocity, we have 

\begin{equation}
U_{ex}=\frac c{\omega _{ce}B_0}\frac{\partial ^2\varphi _e}{\partial
t\partial x}\text{ },  \label{16}
\end{equation}
with 
\[
\varphi _e=\phi +\frac{\hbar ^2}{4m_ee}\nabla ^2N_s-\phi _{p\perp }\text{ }, 
\]
where $\phi _{p\perp }=e\omega _H^2\left\langle \left| E_x\right|
^2\right\rangle /4m_e\omega _{pe}^4$ is the perpendicular (to ${\bf \hat{z}}$%
) component of the UH wave ponderomotive potential. Combining Eqs.\ (14) and
(16) we obtain

\begin{equation}
\left( 1+\lambda _{qe}^4\frac{\partial ^2}{\partial x^2}\nabla ^2\right)
N_s+\left( \frac c{\omega _{ce}B_0}\right) \frac{\partial ^2}{\partial x^2}%
\phi =\frac{\lambda _e^2}{4B_0^2}\frac{\omega _H^2}{\omega _{pe}^2}\frac{%
\partial ^2}{\partial x^2}\left\langle \left| E_x\right| ^2\right\rangle ,
\label{17}
\end{equation}
where $\lambda _{qe}=\left( \hbar ^2/4m_e^2\omega _{ce}^2\right) ^{1/4}$ is
the quantum wavelength of the electrons and $\lambda _e=c/\omega _{pe}$ is
the electron skin depth$.$

In the electrostatic LH field, the ions are unmagnetized and their dynamics
in the quasi-neutrality approximation is governed by Eqs.\ (8) and (9).
Assuming $\omega _{ci}\ll \Omega $ as well as ignoring the ion quantum
diffraction effects, we obtain 

\begin{equation}
\frac{\partial ^2}{\partial t^2}N_s-\frac{c\omega _{ci}}{B_0}\frac{\partial
^2}{\partial x^2}\phi =0\text{ }.  \label{18}
\end{equation}
Eliminating $\phi $ from Eqs.\ (17) and (18), we have 

\begin{equation}
\left( \frac{\partial ^2}{\partial t^2}+\Omega _{LH}^2\right) N_s=\frac{%
\lambda _e^2}{4B_0^2}\frac{\omega _H^2\omega _{LH}^2}{\omega _{pe}^2}\frac{%
\partial ^2}{\partial x^2}\left\langle \left| E_x\right| ^2\right\rangle ,
\label{19}
\end{equation}
which is the driven (by the perpendicular component of the UH ponderomotive
force) electrostatic LH wave equation. Here $\Omega _{LH}=\omega _{LH}\left(
1+\lambda _{qe}^4\partial ^2/\partial x^2\nabla ^2\right) ^{1/2}$, and $\omega
_{LH}=\sqrt{\omega _{ce}\omega _{ci}}$ is the LH resonance frequency. In the
absence of the UH waves, Eq. (19) gives the electrostatic LH wave frequency 

\begin{equation}
\Omega ^2=\omega _{LH}^2\left( 1+\lambda _{qe}^4k_x^2k^2\right) \equiv
\Omega _{LH}^2\text{ ,}  \label{20}
\end{equation}
which exhibits a dispersion due to quantum electron density
correlations. By neglecting the quantum electron wavelength $(\lambda
_{qe}\rightarrow 0),$ we obtain the usual LH resonance frequency.
Equations (5) with $B_1=0,$ (12), and (19) are the desired set for
nonlinearly coupled UH and LH waves in a quantum magnetoplasma.

\subsection{Alfv\'{e}n waves}

Finally, we present the driven Alfv\'{e}n wave equation in a magnetized quantum 
plasma. For this purpose, we use the momentum equations for the inertialess electrons 
and mobile ions, respectively, 

\begin{equation}
0=-e\left( {\bf E}+\frac{{\bf U}_{e1}\times {\bf B}_0}c\right) +\frac{\hbar
^2}{4m_en_{e0}}{\bf \nabla }\nabla ^2n_{e1}-{\bf \hat{x}}\frac{e^2}{4m_e}%
\frac \partial {\partial x}\frac{\omega _H^2}{\omega _{pe}^4}\left\langle
\left| E_x\right| ^2\right\rangle \text{ ,}  \label{21}
\end{equation}
and 

\begin{equation}
m_i\frac{\partial {\bf U}_{i1}}{\partial t}=e\left( {\bf E}+\frac{{\bf U}%
_{i1}\times {\bf B}_0}c\right) \text{ .}  \label{22}
\end{equation}
We have here ignored the quantum diffraction effects and the ponderomotive force
on the ions. Here ${\bf U}_{e1}$ $({\bf U}_{i1})$ is the electron (ion) perturbed 
fluid velocity. Adding Eqs.\ (21) and (22), and introducing the total
current density ${\bf J=}e(n_{i0}{\bf U}_{i1}-n_{e0}{\bf U}_{e1})$ from the
Maxwell equation ${\bf \nabla \times B}_1= 4\pi {\bf J} /c$, and
using $n_{e0}\approx n_{i0},$ we obtain 

\begin{equation}
\frac{\partial {\bf U}_{i1}}{\partial t}=\frac 1{4\pi m_in_{i0}}\left( {\bf %
\nabla \times B}_1\right) \times {\bf B}_0+\frac{\hbar ^2}{4m_em_in_{e0}}%
{\bf \nabla }\nabla ^2n_{e1}-{\bf \hat{x}}\frac{e^2}{4m_em_i}\frac \partial {%
\partial x}\frac{\omega _H^2}{\omega _{pe}^4}\left\langle \left| E_x\right|
^2\right\rangle \text{ ,}  \label{23}
\end{equation}
From (23) we obtain

\begin{equation}
\frac{\partial U_{ix}}{\partial t}=-\frac{V_A^2}{B_0}\frac \partial {%
\partial x}B_1+\frac{\hbar ^2}{4m_em_in_{e0}}\frac \partial {\partial x}%
\nabla ^2n_{e1}-\frac{e^2}{4m_em_i}\frac \partial {\partial x}\frac{\omega
_H^2}{\omega _{pe}^4}\left\langle \left| E_x\right| ^2\right\rangle \text{ ,}
\label{24}
\end{equation}
where $V_A=B_0/\sqrt{4\pi m_in_{i0}}$ is the Alfv\'{e}n speed. By using
the frozen-in field condition $\left( B_1/B_0\right) =\left( n_{i1}/n_{i0}\right) $
in Eq.\ (24) and combining it with  Eq. (8), we have 

\begin{equation}
\left( \frac{\partial ^2}{\partial t^2}-V_a^2\frac{\partial ^2}{\partial x^2}%
\right) N_s=\frac{e^2}{4m_em_i}\frac{\omega _H^2}{\omega _{pe}^4}\frac{%
\partial ^2}{\partial x^2}\left\langle \left| E_x\right| ^2\right\rangle 
\text{ }.  \label{25}
\end{equation}
where $V_a=\left[ V_A^2-\left( \hbar ^2/4m_em_i\right) \nabla ^2\right]
^{1/2}$ is the Alfv\'{e}n speed including the quantum diffraction effects. In
the absence of the UH waves, we have

\begin{equation}
\Omega ^2=k_x^2\left( V_A^2+\frac{\hbar ^2k^2}{4m_im_e}\right) \equiv
k_x^2V_a^2  \label{26}
\end{equation}
Ignoring the electron quantum diffraction effects $\hbar \rightarrow 0,$ we obtain 
from (26) the frequency of the usual Alfv\'{e}n waves in an electron ion plasma. 
Equations (5) and (25) are the desired set for investigating the parametric interactions 
between the UH and Alfv\'{e}n waves in a quantum magnetoplasma. 

In the following, we shall study
the decay and modulational instabilities of an UH wave involving the IC, LH, and Alfv\'en waves in
a quantum magnetoplasma.

\section{Nonlinear Dispersion Relations and Growth Rates}

In this section, we shall derive the nonlinear dispersion relations 
for three-wave decay and modulational instabilities.
  
\subsection{Coupling of UH and IC waves}

To derive the nonlinear dispersion relation for parametric instabilities in a quantum 
magnetoplasma, we write the UH electric field as the sum of the pump
wave and the upper and lower UH sideband fields. The latter arise due the coupling of the 
pump $E_{x0}$ exp$(i{\bf k}_0\cdot {\bf r}-i\omega _0t)$ $+c.c.$ with
low-frequency IC, LH and Alfv\'enic perturbations. Specifically, the high-frequency 
UH pump $(\omega _0,{\bf k}_0)$ interacts with the low-frequency electrostatic IC waves $%
(\Omega ,{\bf k})$ having $N_s=\hat{N}_s$ exp$(i{\bf k}\cdot {\bf %
r}-i\Omega t),$ and produces two UH sidebands $E_{x\pm }$ exp$(i{\bf k}_{\pm }\cdot 
{\bf r}-i\omega _{\pm }t)$, with frequencies $\omega _{\pm }=\Omega \pm
\omega _0$ and wavenumbers ${\bf k}_{\pm }={\bf k\pm k}_0.$ By using the Fourier
transformation, and matching phasors, we obtain from Eq. (5) with $%
B_1=0,$ and Eq.\ (12)  

\begin{equation}
D_{\pm }E_{x\pm }=\omega _{pe}^2\hat{N}_sE_{x0\pm}  , \label{27}
\end{equation}
where $E_{x0+} = E_{x0}$ and $E_{x0-} = E_{x0}^*$, and 
\begin{equation}
\left( \Omega ^2-\Omega _{IC}^2\right) \hat{N}_s=\frac{k_x^2}{16\pi n_{e0}m_i%
}(E_{x0}^{*}E_{x+}+E_{x0}E_{x-})\text{ ,}  \label{28}
\end{equation}
where the asterisk denotes the complex conjugate. The upper and lower sidebands can 
be written as

\begin{equation}
D_{\pm }=\omega _{\pm }^2-\omega _H^2-\frac{\hbar ^2}{4m_e^2}k_{x\pm
}^2k_{\pm }^2\text{ .}  \label{29}
\end{equation}
For $\Omega \ll \omega _{0,}$ (29) reduces to

\begin{equation}
D_{\pm }=\pm 2\omega _0\left( \Omega \mp \Delta -\delta \right) \text{ ,}
\label{30}
\end{equation}
where $\omega _0=\sqrt{\omega _H^2+(\hbar ^2/4m_e^2)\text{ }k_{x0}^2k_0^2}$
is the UH wave frequency modified by the quantum effects, $\Delta =\left(
\hbar ^2/8m_e^2\omega _0\right) (k_x^2k_0^2+k_{x0}^2k^2+k_x^2k^2+4k_{x0}k_x%
{\bf k\cdot k}_0),$ and $\delta =\left( \hbar ^2/4m_e^2\omega _0\right)
\left\{ k_{x0}k_x\left( k^2+k_0^2\right) +{\bf k\cdot k}_0\left(
k_x^2+k_{x0}^2\right) \right\} $ are the frequency shifts arising from the
nonlinear coupling between the  UH and IC waves. Eliminating $E_{x+}$ and $%
E_{x-}$ from Eq. (27) and Eq. (28), we have

\begin{equation}
\Omega ^2-\Omega _{IC}^2=\frac{\omega _{pe}^2\text{ }k_x^2\text{ }\left|
E_{x0}\right| ^2}{16\pi n_{e0}m_i}\sum_{+,-}\frac 1{D_{\pm }} \, .  \label{31}
\end{equation}
Equation (31) is the dispersion relation for parametrically 
coupled UH and IC waves in a quantum magnetoplasma.

For three-wave decay interaction, we consider the
lower sideband $D_{-}$ to be resonant, while the upper sideband $D_{+}$ is
assumed off-resonant. We then obtain from (31)

\begin{equation}
\left( \Omega ^2-\Omega _{IC}^2\right) \left( \Omega +\Delta -\delta \right)
=-\frac{\omega _{pe}^2\text{ }k_x^2\text{ }\left| E_{x0}\right| ^2}{32\pi
n_{e0}m_i\omega _0}\text{ }.  \label{32}
\end{equation}
Letting $\Omega =\Omega _{IC}+i\gamma _{IC}$ and $\Omega =\delta -\Delta
+i\gamma _{IC}$ with $\Omega _{IC}\sim \delta -\Delta ,$ we obtain from (32)
for $\gamma _{IC}\ll \Omega _{IC},$ the growth rate 

\begin{equation}
\gamma _{IC}\simeq \frac{\omega _{pe}\text{ }k_x\text{ }\left| E_{x0}\right| 
}{8\sqrt{\pi n_{e0}m_i\omega _0\Omega _{IC}}}  \label{33}
\end{equation}

For the modulational instability, both the lower and upper sidebands $D_{\pm
}$ are resonant. Thus, Eq. (31) gives

\begin{equation}
\left( \Omega ^2-\Omega _{IC}^2\right) \left[ \left( \Omega -\delta \right)
^2-\Delta ^2\right] =\frac{\omega _{pe}^2\text{ }k_x^2\text{ }\left|
E_{x0}\right| ^2}{16\pi n_{e0}m_i\omega _0}\Delta \text{ }.  \label{34}
\end{equation}
Assuming $\Omega \gg \delta ,$ we obtain 

\begin{equation}
\Omega ^4-\left( \Delta ^2+\Omega _{IC}^2\right) \Omega ^2+\Delta ^2\Omega
_{IC}^2-\frac{\omega _{pe}^2\text{ }k_x^2\text{ }\left| E_{x0}\right| ^2}{%
16\pi n_{e0}m_i\omega _0}\Delta =0\text{ .}  \label{35}
\end{equation}
The solutions of Eq. (35) are 


\begin{equation}
\Omega ^2=\frac 12\left[ \Delta ^2+\Omega _{IC}^2\pm \sqrt{\left( \Omega
_{IC}^2-\Delta ^2\right) ^2 + \Omega _{m1}^4}\right] \text{ },  \label{37}
\end{equation}
where

\begin{equation}
\Omega _{m1}=\left( \frac{\omega _{pe}^2k_x^2\Delta }{%
4\pi n_{e0}m_i\omega _0}\right) ^{1/4}\left| E_{x0}\right|^{1/2}  \, .  \label{38}
\end{equation}
The growth rate of the modulational instability is

\begin{equation}
  \gamma_{m1} = \left( \frac{\omega _{pe}^2k_x^2|\Delta| }{%
16\pi n_{e0}m_i\omega _0}\right) ^{1/4}\left| E_{x0}\right|^{1/2}  \, .
\end{equation}

\subsection{Coupling of UH and LH waves}

In this case, the UH pump wave interacts with the low-frequency
electrostatic LH waves $({\bf k,}\Omega )$.  By using Fourier 
transformations and matching phasors, we obtain from Eq. (5) with $B_1=0$, 
and Eq. (19)  

\begin{equation}
D_{\pm }E_{x\pm }=\omega _{pe}^2\hat{N}_s E_{x0\pm} 
\, ,  \label{39}
\end{equation}
and 

\begin{equation}
\left( \Omega ^2-\Omega _{LH}^2\right) \hat{N}_s=\frac{k_x^2\lambda
_e^2\omega _H^2\omega _{LH}^2}{4B_0^2\omega _{pe}^2}%
(E_{x0}^{*}E_{x+}+E_{x0}E_{x-})\text{ ,}  \label{40}
\end{equation}
where $D_{\pm }=\pm 2\omega _0\left( \Omega \mp \Delta -\delta \right) $ for 
$\Omega \ll \omega _0$, $\Delta =\left( \hbar ^2/8m_e^2\omega _0\right)
(k_x^2k_0^2+k_{x0}^2k^2+k_x^2k^2+4k_{x0}k_x{\bf k\cdot k}_0),$ and $\delta
=\left( \hbar ^2/4m_e^2\omega _0\right) \left\{ k_{x0}k_x\left(
k^2+k_0^2\right) +{\bf k\cdot k}_0\left( k_x^2+k_{x0}^2\right) \right\} $ are
the frequency shifts arising from the nonlinear coupling of the UH waves with the
LH waves. Inserting the expressions for $E_{x+}$ and $E_{x-}$ from Eq. (39) into
Eq. (40), we find the nonlinear dispersion relation

\begin{equation}
\Omega ^2-\Omega _{LH}^2=\frac{k_x^2\lambda _e^2\omega _H^2\omega _{LH}^2%
\text{ }\left| E_{x0}\right| ^2}{4B_0^2}\sum_{+,-}\frac 1{D_{\pm }}%
 \text{ }.  \label{41}
\end{equation}

Since for three-wave decay interactions, the lower and upper sidebands $D_{-}$ $%
(D_{+})$ are resonant (off-resonant), we obtain from (41)

\begin{equation}
\left( \Omega ^2-\Omega _{LH}^2\right) \left( \Omega +\Delta -\delta \right)
=-\frac{k_x^2\lambda _e^2\omega _H^2\omega _{LH}^2\text{ }\left|
E_{x0}\right| ^2}{8B_0^2\omega _0}\text{ }.  \label{42}
\end{equation}
Letting $\Omega =\Omega _{LH}+i\gamma _{LH}$ and $\Omega =\delta -\Delta
+i\gamma _{LH},$ with $\Omega _{LH}\sim \delta -\Delta ,$ we obtain the growth rate 
from Eq. (42), under the approximation $\gamma _{LH}\ll \Omega _{LH}$, 

\begin{equation}
\gamma _{LH}\simeq \frac{k_x\lambda _e\omega _H\omega _{LH}\text{ }\left|
E_{x0}\right| }{4B_0\sqrt{\omega _0\Omega _{LH}}}\text{ .}  \label{43}
\end{equation}

Since for the modulational instability, both the sidebands $D_{\pm }$ are
resonant, we have from (41)

\begin{equation}
\left( \Omega ^2-\Omega _{LH}^2\right) \left[ \left( \Omega -\delta \right)
^2-\Delta ^2\right] =\frac{k_x^2\lambda _e^2\omega _H^2\omega _{LH}^2\text{ }%
\left| E_{x0}\right| ^2}{4B_0^2\omega _0}\Delta \text{ }.  \label{44}
\end{equation}
Simplifying Eq. (44) for $\Omega \gg \delta ,$ we have $_{}$%

\begin{equation}
\Omega ^4-\left( \Delta ^2+\Omega _{LH}^2\right) \Omega ^2+\Delta ^2\Omega
_{LH}^2-\frac{k_x^2\lambda _e^2\omega _H^2\omega _{LH}^2\text{ }\left|
E_{x0}\right| ^2}{4B_0^2\omega _0}\Delta =0\text{ .}  \label{45}
\end{equation}
Equation (45) admits the solutions

\begin{equation}
\Omega ^2=\frac 12\left( \Delta ^2+\Omega _{LH}^2\right) \pm \frac 12\left[
\left( \Delta ^2-\Omega _{LH}^2\right) ^2 + \Omega _{m2}^4\right] ^{1/2}, 
\label{46}
\end{equation}
where 

\begin{equation}
\Omega _{m2} =\left( \frac{k_x^2\lambda _e^2\omega _H^2\omega _{LH}^2\text{ }%
}{B_0^2\omega _0}\Delta \right) ^{1/4}\left| E_{x0}\right|^{1/2}  \, .  \label{47}
\end{equation}

\subsection{Coupling of UH and Alfv\'{e}n waves}

Finally, we consider the nonlinear interaction of the UH pump wave with 
Alfv\'{e}n waves $(\Omega, {\bf k})$. We follow the same procedure as described above,
and obtain 

\begin{equation}
D_{\pm }E_{x\pm }= (\omega _{pe}^2+ 2 \omega_{ce}^2) \hat{N}_s
E_{x0\pm} \text{ ,}  \label{48}
\end{equation}
and 

\begin{equation}
\left( \Omega ^2-k_x^2V_a^2\right) \hat{N}_s=\frac{e^2k_x^2}{4m_em_i}\frac{%
\omega _H^2}{\omega _{pe}^4}\left( E_{x0}^{*}E_{x+}+E_{x0}E_{x-}\right) 
\text{ ,}  \label{49}
\end{equation}
where $D_{\pm }=\pm 2\omega _0\left( \Omega \mp \Delta -\delta \right) $
with $\Delta =\left( \hbar ^2/8m_e^2\omega _0\right)
(k_x^2k_0^2+k_{x0}^2k^2+k_x^2k^2+4k_{x0}k_x{\bf k\cdot k}_0)$ and $\delta
=\left( \hbar ^2/4m_e^2\omega _0\right) \left\{ k_{x0}k_x\left(
k^2+k_0^2\right) +{\bf k\cdot k}_0\left( k_x^2+k_{x0}^2\right) \right\} $ are
the frequency shifts arising from the nonlinear coupling of the UH waves with
the Alfv\'{e}n waves. Combining Eqs.\ (48) and (49), we have the
nonlinear dispersion relation 

\begin{equation}
\Omega ^2-k_x^2V_a^2=\frac{e^2k_x^2}{4m_em_i}\frac{(\omega_{pe}^2 + 2 \omega_{ce}^2)\omega _H^2}
{\omega _{pe}^4}\left| E_{x0}\right| ^2\sum_{+,-}\frac 1{D_{\pm }} 
\text{ }.  \label{50}
\end{equation}
Proceeding as before, Eq. (50) yields, respectively,  

\begin{equation}
\gamma _{AL}\simeq \frac{e(\omega_{pe}^2 + 2 \omega_{ce}^2)^{1/2}\omega _H\left| E_{x0}\right| }
{4\omega _{pe}^2} \sqrt{\frac{k_x\text{ }}{m_em_i\omega _0V_a}}   \label{51}
\end{equation}
and 
\begin{equation}
\gamma _{m3}=\left( \frac{e^2k_x^2 (\omega_{pe}^2 + 2 \omega_{ce}^2)\omega _H^2}
{4m_e m_i\omega _0\omega _{pe}^4}|\Delta| \right) ^{1/4}\left| E_{x0}\right|^{1/2} \,   \label{52}
\end{equation}
for the growth rates of the three-wave decay and modulational instabilities in quantum
magnetoplasmas when the  UH and Alfv\'{e}n waves are nonlinearly coupled.

\section{Summary}

In summary, we have considered the nonlinear couplings between UH,
IC, LH, and Alfv\'{e}n waves in a quantum magnetoplasma. We have derived 
the governing nonlinear equations and the appropriate dispersion relations 
by employing the one-dimensional quantum magnetohydrodynamical equations. It is 
found that the wave dispersion is due to the quantum correction
arising from the strong electron density correlations at quantum scales.
The dispersion relations have been analyzed analytically to obtain the growth
rates for both the decay and modulational instabilities involving 
dispersive IC, LH and Alfv\'en waves. Since the frequencies of the
latter are significantly modified due to the quantum corrections, the 
growth rates are accordingly affected in quantum magnetoplasmas.
The present results can be important for diagnostic purposes in
magnetized quantum systems, such as those in dense astrophysical 
objects, intense laser-matter experiments, and in dense semiconductor
devices in an external magnetic field. 

{\bf Acknowledgments:} S. A. acknowledges financial support from the
Deutscher Akademischer Austauschdienst.

\end{document}